\documentclass[prl,twocolumn,showpacs]{revtex4}
\usepackage{graphicx}
\usepackage{hyperref}
\usepackage{color}

\def\hhref#1{\href{http://arxiv.org/abs/#1}{arXiv:#1}} 

\newcommand{\be}{\begin{equation}}
\newcommand{\ee}{\end{equation}}
\newcommand{\bea}{\begin{eqnarray}}
\newcommand{\eea}{\end{eqnarray}}

\begin{document}


\title{ Thermodynamics of photons on fractals}
\author{Eric~Akkermans$^{1}$, Gerald~V.~Dunne$^{2}$ and Alexander~Teplyaev$^{3}$}
\affiliation{$^{1}$ Department of Physics, Technion Israel Institute of Technology,
  32000 Haifa, Israel   \\ $^{2}$Department  of Physics, University of Connecticut, Storrs, CT 06269 \\  $^{3}$Department  of Mathematics, University of Connecticut, Storrs, CT 06269}

\begin{abstract}
A thermodynamical treatment of a massless scalar field (a "photon") confined to a fractal spatial manifold leads to an equation of state relating pressure to internal energy, $P V_s=U/d_s$, where $d_s$  is the spectral dimension and  $V_s$ defines the  "spectral volume". For regular manifolds, $V_s$ coincides with the usual geometric spatial volume, but on a fractal this is not necessarily the case. This is further evidence that on a fractal, momentum space can have a different dimension than position space.
Our analysis also provides a natural definition of the vacuum (Casimir) energy of a fractal. We suggest ways that these unusual properties might be probed experimentally.

\end{abstract}



\maketitle


Among one of the more unusual properties of spatial fractals is that in some not yet completely understood sense the dimension of momentum space can be different from the dimension of position space. Even the notion of a (Fourier) mode decomposition on a fractal is not yet understood mathematically \cite{kigami,lapidus,uncertainty}. At first sight, this would appear to render problematical the conventional formulation of thermodynamics and statistical mechanics in terms of phase space cells \cite{lifshitz}. Indeed, the quantization of the electromagnetic field and resulting notion of a photon appears inherently based on the Fourier decomposition of the field as a superposition of an infinite number of modes and their quantization as bosonic quantum harmonic oscillators. 
The very idea of mode decomposition is based on the assumption that it is possible to label and count the modes by quantizing 
momentum space
into elementary cells of volume $L^{-d}$, where the spatial volume $V= L^d$ is taken to be a large hypercube in $d$ space dimensions. 

While such a mode-based approach is not available at present for fractals, some results are known on fractals concerning the spectrum of the Laplacian, encoded in spectral functions such as the heat kernel trace and  zeta function \cite{kigami,lapidus,kl,adt1,strichartz}. We argue  this is sufficient information to define a consistent thermodynamic limit, leading to an equation of state $P V_s=U/d_s$, in terms of a "spectral volume" $V_s$ [defined below] and the spectral dimension, $d_s$ \cite{orbach}.
On a regular manifold, the spectral and geometric volume coincide, as do $d_s$ and the geometric dimension 
$d$, but on a fractal this is not necessarily true.

The idea of using thermodynamics to probe the geometrical structure of objects goes back to H. Lorentz, who in 1910 addressed the original "Can one hear the shape of a drum?" question \cite{kac}, asking why the Jeans law, $dE =V(T/\pi^2c^3) \omega^2 d\omega$, only depends on the volume occupied by  the black-body, and not on its shape. Soon after, H. Weyl found the geometric meaning of the large eigenvalues of differential operators \cite{weyl-proof,weyl-review}, and it turns out that this is what is probed in the thermodynamic large volume and high temperature limit. 
Physically, at thermodynamical equilibrium the photons  probe the geometric structure of the spatial manifold in which they are confined; the first geometric property we learn  is the volume. We are also motivated by suggestions of fractal structure in models of quantum gravity \cite{zam}.

Our main technical result is an expression for the partition function $\mathcal Z (T, V_s)$ of quantum radiation at thermodynamic equilibrium at temperature $T$, on a fractal:
\begin{equation}
\ln \mathcal Z = y\,  {\mathcal F}_{\rm per} \left( \ln y \right)  \qquad, \quad y\equiv  V_s/(\beta \hbar c)^{d_s} \, \ .
\label{part}
\end{equation}
Here, $\beta\equiv 1/(k_B T)$, $k_B$ is Boltzmann's constant, and 
${\mathcal F}_{\rm per}$ is a periodic function of $\ln \left(y\right)$, whose period depends on the characteristics of the fractal.

We first review the usual relation between the partition function and the effective Lagrangian for a massless scalar field in Euclidean space-time, emphasizing the importance of the asymptotic limit of large volumes or correspondingly large temperatures, and its relation to the Weyl expansion \cite{dowker}. In this approach, the mode decomposition is encoded in the heat kernel trace for the Laplace operator, so we can then generalize to fractals.

The partition function can be written as:
\begin{equation}
\ln \mathcal Z (T,V) =-\frac{1}{2} \ln \mbox{Det}_{M \times \cal M} \left( \partial ^2 / \partial t^2 + c^2 \Delta \right)
\label{part2}
\end{equation}
$M$ is a circle of circumference $\hbar \beta$ in the coordinate $t$,
$\Delta$ is (minus) the Laplacian defined on the spatial manifold $\cal M$, and $c$ is the speed of light. While (\ref{part2}) is independent of any specific representation,  it is usually derived using a mode decomposition of the field. 
Recall that the spectral partition function for a single bosonic oscillator of frequency $\omega$ is
\be
\ln Z (T,\omega)  = - { \beta \hbar \omega \over 2} - \ln \left( 1 - e^{- \beta \hbar \omega} \right) \, \ .
\label{eq3}
\ee
By standard manipulations \footnote{First note the integral representation: $\frac{1}{n}e^{-\beta \hbar \omega n}=\frac{\beta \hbar}{\sqrt{4\pi}}\int_0^\infty \frac{d\tau}{\tau^{3/2}} \,e^{-\omega^2 \tau}\, e^{-(n\beta \hbar)^2/(4\tau)}$;  then the Poisson summation relation: $\sqrt{ \frac{\pi}{t}} \sum_{n = - \infty}^{+ \infty} e^{- {\pi^2 n^2\over t}}=
 \sum_{n = - \infty}^{+ \infty} e^{-  t n^2 }$.}, we obtain
 \be
\ln Z (T, \omega)  =\frac{1}{2} \int_0^\infty {d \tau \over \tau} e^{- \omega^2 \tau}  \sum_{n = - \infty}^{+ \infty} e^{-  ({2 \pi n\over \hbar \beta})^2 \tau} \, \ .
\label{eq5}
\ee
Using the  Kubo, Martin and Schwinger (KMS) condition \cite{kms} for thermodynamic equilibrium of a bosonic field $\phi$ at temperature $T$, we recognize the second factor in the integrand as a sum over Matsubara frequencies $(\frac{2 \pi n}{\hbar \beta})^2$, corresponding to the spectrum of the operator $\partial^2/ \partial t^2$ with periodic boundary conditions $\phi (t + \hbar \beta ) = \phi (t)$.
Identifiying $\omega^2=c^2 k^2$ with the eigenvalues of $c^2\Delta$, and tracing over all modes  \footnote{Note: $\ln \mathcal O=-\int_0^\infty \frac{d\tau}{\tau}e^{-\mathcal O\, \tau}$, and $\ln {\rm Det}\, \mathcal O={\rm Tr}\,\ln  \mathcal O$.}, we recover expression (\ref{part2}): 
\be
\ln \mathcal Z (T,V)  = {1 \over 2} \, \int_0^\infty {d \tau \over \tau}  \mbox{Tr}_{\cal M}  \left( e^{- \tau c^2 \Delta} \right)\,  \mbox{Tr}_{M} \left( e^{- \tau \partial_0 ^2} \right) \, \ .
\label{eq6}
\ee
All  information about the spatial manifold ${\mathcal M}$, including possible dependence on its volume $V$, is contained in the heat kernel trace factor. It is convenient to rescale $\tau$:
\be
\ln \mathcal Z (T,V)  = \frac{1}{2} \int_0^\infty {d \tau \over \tau} f ( \tau ) \, {\cal K}_{\cal M} (L_\beta^2\,  \tau )
\label{eq8}
\ee
defining the "photon" thermal  wavelength, $L_\beta\equiv \beta\hbar c$,   
the dimensionless function $f (\tau) = \sum_{n = - \infty}^{+ \infty} e^{- (2\pi n)^2 \tau}$, and $\mathcal K_{\mathcal M}$
 as the dimensionless heat kernel trace of the Laplace operator on the manifold $\mathcal M$ 
\be
\mathcal K_{\mathcal M} (L_\beta^2 \, \tau) \equiv {\rm Tr}_{\mathcal M} e^{- L_\beta^2 \, \tau\,{\Delta}} \, \ .
\label{hk}
\ee
The heat kernel contains an implicit length scale $L$, characteristic of  the geometry of the manifold $\mathcal M$, that enables us to define a dimensionless Laplacian $\tilde{\Delta}\equiv L^2\, \Delta$ [recall that 
the eigenvalues of $\Delta$ have dimensions of $1/{\rm length}^2$]. 
Results in the mathematical literature refer to the dimensionless Laplacian $\tilde{\Delta}$.

As an illustration, consider black-body radiation  in a 
large hypercube of volume $V = L^d$, in $d$ space dimensions. The corresponding heat kernel ${\cal K}_{V} (\tau )$ is obtained from the mode decomposition of the field, $\omega = c | {\bf k} | = c\, 2 \pi |{\bf n}|/V^{1/d}$, where $\bf n$ is a $d$-dimension integer-valued vector obtained by counting the elementary momentum-space cells of volume $(2\pi)^d/V$: 
\be
 {\cal K}_{V } (\tau ) = \sum_{\bf n} e^{- (2\pi L_\beta V^{- 1/d} )^2  {\bf n}^2\,\tau} \, \ . 
 \label{cube}
 \ee
 It is clear that
 the partition function has the form
 \be
\ln \mathcal Z (T,V)  = Q \left(x \right) \qquad, \quad x\equiv  L_\beta/V^{1/d}
\label{zeuclide}
\ee
where the precise form of the function $Q$ is determined by the heat kernel trace in (\ref{hk}).
Standard thermodynamic quantities and relations follow immediately from (\ref{zeuclide}). For example, the equation of state, $P V = U/d$, relating the  internal energy $U$ of the radiation to the pressure $P$ and the volume $V$, follows immediately from
\be
U = - {\partial \over \partial \beta} \ln \mathcal Z (T,V) = - \left( {d Q \over d x} \right)  \hbar \, c \,  V^{- 1/d}
\label{eqU}
\ee
and 
\be
P = {1 \over \beta} \left( {\partial \ln \mathcal Z \over \partial V} \right)_T = - \left( {d Q \over d x} \right) {\hbar c  V^{- 1/d} \over V \, d} \, \ .
\label{pressure}
\ee
The Stefan-Boltzmann expression for the internal energy $U$ is a consequence of the equation of state and the thermodynamic relation $
 \left( {\partial U \over \partial V} \right)_T = T  \left( {\partial P \over \partial T} \right)_V - P$, while noticing from (\ref{pressure}) that $P$ depends on $T$ only, in the thermodynamic limit. We then obtain $U = a V T^{d+1}$ where $a$ is a constant to be determined from (\ref{eq8}). 
Also, the entropy, 
 \be S = k_B U/T + k_B \ln \mathcal Z (T,V) = d \,  x \, Q' (x) + Q(x) \, \ , \label{eqs} \ee 
depends only on the dimensionless parameter $x$, so that an adiabatic (isentropic) expansion of the radiation occurs at  constant values of the product $V T^d $. 
 
 For black-body radiation in manifolds of complicated shape, it is difficult to make an explicit mode decomposition and  find an explicit expression like (\ref{cube}) for the heat kernel trace. 
However, (\ref{zeuclide}) continues to hold. We
 can learn about the thermodynamic [large volume] limit from the Weyl expansion of the heat kernel trace, described below. Note that the large volume limit corresponds to $V \gg  L_\beta ^d$, which is a high temperature limit $ k_B T \gg \hbar c /V^{1/d}$. This probes the small $\tau$ behavior of $\mathcal K_{\mathcal M}(\tau)$ for which a general asymptotic expansion is known:
 \be
 {\cal K}_{\mathcal M} (\tau )  \sim {V \over (4 \pi L_\beta ^2 \tau )^{d/2}} - \alpha {S \over (4 \pi L_\beta ^2 \tau )^{\frac{d-1}{2}}} + \cdots 
 \label{weyl}
 \ee
where $V$ is the radiation cavity volume, $S$ is the surface area, and $\alpha$ is a numerical constant depending on the boundary conditions. Higher-order terms in the expansion characterize other geometric and topological properties such as the curvature of the surface \cite{weyl-review,baltes}. Keeping only the dominant  volume term in (\ref{weyl}), expression (\ref{eq8})  leads immediately to the familiar thermodynamic expressions \cite{lifshitz}: $\ln\mathcal Z=(V/L_\beta^d) \zeta_R(d+1)\Gamma\left(\frac{d+1}{2}\right)/\pi^{(d+1)/2}$, $P=(k_B T /L_\beta^d) \zeta_R(d+1)\Gamma\left(\frac{d+1}{2}\right)/\pi^{(d+1)/2}$. 
Away from the thermodynamic  limit, subdominant terms in (\ref{weyl}) 
lead to corrections that depend on the exact geometry of the volume enclosing the radiation, but the equation of state $PV=U/d$ is always valid.
This formulation in terms of heat kernels makes it clear that the Weyl expansion is directly related to the thermodynamic limit of a black-body radiation system, so we can use the leading term as a {\it definition} of the volume probed by the photons as they attain thermal equilibrium.
\begin{figure}[htb]
\includegraphics[scale=0.22]{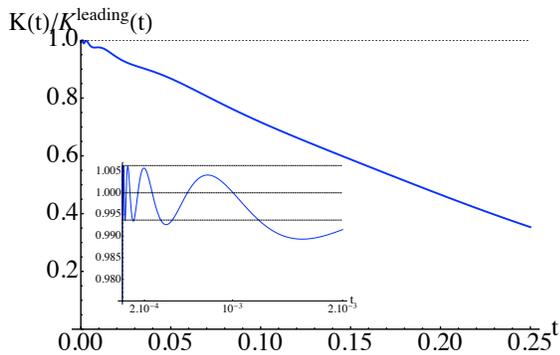}
\caption{Typical small $\tau$ behavior of a fractal heat kernel trace. There are small oscillations [see inset], as a function of $\ln \tau$, about the leading behavior $\sim 1/\tau^{d_s/2}$.}
\end{figure}

We now turn to fractals, and make use of recent progress concerning the heat kernel trace on fractals \cite{kigami,lapidus,kl,adt1,strichartz}.
While it is familiar that fractals have a nontrivial fractal (Hausdorff) dimension $d_h$, arising from their self-similar spatial structure, it is perhaps less well-known that fractals are characterized by another dimension, known as the "spectral dimension" $d_s$,  which need not be equal to $d_h$ \cite{orbach}. Indeed, diffusion on fractals is  "anomalous" in the sense that the Einstein Brownian motion relation, $\langle r^2 (t) \rangle \propto  t$, for long enough times, is replaced on a fractal by 
$ \langle r^2 (t) \rangle \propto t^{2/d_w} $, where $d_w=2d_h/d_s$ is called the ``anomalous random walk dimension'' \cite{orbach}. 
 For many fractals, including well-studied cases such as the Sierpinski gaskets or diamonds, the small $\tau$ limit of the heat kernel trace is more complicated than the Weyl expansion expression in (\ref{weyl}). The  heat kernel trace $\mathcal K_{\mathcal F}(\tau)$, on a fractal $\mathcal F$, for the dimensionless Laplacian $\tilde{\Delta}$, behaves as $\mathcal K_{\mathcal F}(\tau) = \tau^{- d_s / 2} \, F(\tau)$, where $F$ is a periodic function of $\ln \tau$ \cite{kl}. 
The typical small $\tau$ behavior is shown in Figure 1. 
The leading small $\tau$ behavior, $\mathcal K_{\mathcal F}(\tau) \sim\tau^{- d_s / 2}$, is determined by the spectral dimension $d_s$, but there are also oscillations around this leading behavior, and these oscillations are typically very small.
Explicit expressions for this oscillatory behavior are given in \cite{adt1} for diamond fractals, and numerical results for the Sierpinski gasket appear in \cite{strichartz}. 
Restoring units for the Laplacian, there exists a characteristic length $L_s$ (a "spectral length") such that,
\be
\mathcal K_{\mathcal F}(\tau) =  \left( {L_s ^2 \over L_\beta ^2 \tau}  \right)^{d_s / 2} F\left( L_\beta ^2  \tau / L_s ^2 \right) \, \ .
 \label{hkfractal}
\ee
Inserting this expression into (\ref{eq8}) leads to (\ref{part}) for the partition function $\ln \mathcal Z$. 

The first important conclusion from this result is that the thermodynamic volume is determined  by the spectral dimension $d_s$ rather than the fractal dimension $d_h$. So, we  define the thermodynamic volume of the fractal to be the "spectral volume": $V_s= \mbox{b} \, L_s ^{d_s}$, where $\mbox{b}$  is a numerical coefficient that we specify in detail below. This implies that the thermodynamic equation of state on a  fractal is
\be
P \, V_s \, = \, {1 \over d_s } \, U
\label{eqfract}
\ee
and according to (\ref{eqs}), adiabatic (isentropic) transformation on fractals occur at constant values of the product $V_s \, T^{d_s} $.
The second important conclusion is that the actual expressions for pressure $P$, internal energy $U$, etc... will be modified on a fractal, not only by the appearance of the spectral dimension and spectral volume, but also by the appearance of oscillatory terms arising from the behavior of the log periodic function $F$ in (\ref{hkfractal}).

In order to make this more explicit, and to give the actual numerical coefficient appearing in the spectral volume, it is convenient to make a Mellin transform to convert  our expressions from the heat kernel trace (\ref{hk}) to the associated (dimensionless)  zeta function \cite{hawking}:
\be
 \zeta_{\mathcal M}(s) \equiv {\rm Tr}_{\mathcal M}\frac{1}{(\tilde{\Delta})^s} =\frac{1}{\Gamma (s)} \int_{0}^{\infty} \frac{d\tau}{\tau} \
\tau^{s}\, {\rm Tr}\,e^{-\tilde{\Delta}\, \tau}
\label{zeta1} \ee
with inverse  ($C$ is the usual inverse Mellin contour)
\be
{\cal K}_{\cal M}(L_\beta ^2 \tau)=\frac{1}{2\pi i}\int_C ds \,\left(\frac{L}{L_\beta}\right)^{2s}\zeta_{\cal M} (s)\, \Gamma(s) \, \tau^{-s} \, \ .
\label{inverse}
\ee
Straightforward manipulations show that the partition function can now be expressed as 
\bea
\ln \mathcal Z(T, V)&=&-\frac{1}{2}\left(\frac{L_\beta}{L}\right)\,\zeta_{\mathcal M}\left(-\frac{1}{2}\right)\nonumber\\
&&\hskip -2cm +\frac{1}{\pi i}\int_C \left(\frac{L}{L_\beta}\right)^{2s}\, \Gamma(2s)\, \zeta_{R}(2s+1)\, \zeta_{\mathcal M}(s) \, ds
\label{zetaform}
\eea
The first term gives the standard zero temperature ``vacuum energy" contribution \cite{dowker}, proportional to $\zeta_{\mathcal M}\left(-\frac{1}{2}\right)$,
which has recently been generalized to quantum graphs  \cite{harrison}. On a fractal ${\mathcal F}$, our thermodynamical analysis gives a simple expression for the Casimir energy: 
\be
E_0 = { 1 \over 2} {\hbar c \over L_s} \zeta_{\mathcal F} \left( -\frac {1}{ 2} \right) \, \ .
\label{casimir}
\ee 
The second term in (\ref{zetaform}) encodes  finite temperature corrections to the internal energy $U$; the Riemann zeta  factor $\zeta_{R}(2s+1)$ arises from the sum over Matsubara modes.

On a regular manifold $\mathcal M$, the zeta function $\zeta_{\mathcal M}(s)$ is a meromorphic function in the complex plane with simple poles on the real axis. The pole with largest real part is at $s=d/2$, and gives the leading term in the Weyl expansion (\ref{weyl}). Thus, 
we find a natural thermodynamical definition of  the volume in terms of the residue of $\zeta_{\mathcal M}(s)$ at this pole: $V=(4\pi)^{d/2}\Gamma(d/2) L^d\, {\rm Res}(d/2)$. Depending on the form of the manifold, and the boundary conditions, there may be other poles on the real axis, and these determine the subleading terms in (\ref{weyl}). 
\begin{figure}[htb]
\includegraphics[scale=0.23]{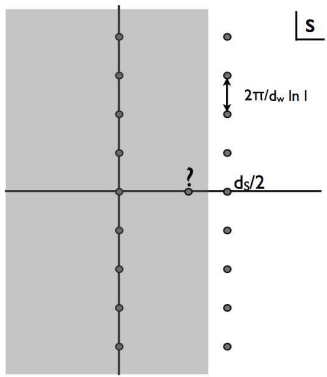}
\caption{The generic pole structure of the zeta function on a symmetric fractal \cite{teplyaev,derfel}. The dominant behavior comes from the real pole at $s=d_s/2$, and oscillations from the tower of complex poles in (\ref{poles}). There may be other poles in the shaded region to the left of this tower, but these do not contribute to the thermodynamic limit. On the Sierpinksi gasket, there is another tower of complex poles on the imaginary axis.}
\end{figure}
When the spatial manifold $\mathcal M$ is a fractal, the situation is radically different 
\cite{kl,lapidus,teplyaev,derfel}: the zeta function of the (dimensionless) Laplacian $\tilde{\Delta}$ has also been shown to be meromorphic in the complex $s$ plane, with simple poles, but now the poles can be complex, as illustrated in Figure 2. The pole with the largest real part lies on the real axis at $s=d_s/2$, where $d_s$ is the spectral dimension. This pole leads to the dominant behavior, shown in the pre-factor in (\ref{hkfractal}), and provides the definition of the spectral volume:
\be
V_s\equiv \left(4\pi\right)^{d_s /2}\Gamma\left(\frac{d_s}{2}\right)L_s ^{d_s}\, {\rm Res}\left(\frac{d_s}{2}\right) \, \ .
\label{vs}
\ee
In addition, there is an infinite tower of complex poles with the same real part:
\be
s_m = \frac{d_s}{2}+ {2 i \pi m \over d_w \ln l}\quad , \quad m\in {\bf Z}
\label{poles}
\ee
where $l$ denotes the number of pieces the basic fractal unit is split into in each iteration \cite{teplyaev,derfel,adt1}.
These complex poles are the origin of the log periodic oscillations in (\ref{hkfractal}), and have been identified with complex dimensions for fractals \cite{kigami,teplyaev}. Explicit expressions for the zeta function on diamond fractals lead to simple expressions for these poles and their residues \cite{adt1}. Integral representations have been given in \cite{derfel} for the zeta function on other fractals, including the Sierpinski gaskets.
In \cite{adt1} it was argued that the origin of these complex poles is the appearance of exponential degeneracy factors, rather than power-law degeneracy factors that appear for regular manifolds. 

To conclude, we have shown that the thermodynamical volume of a fractal is determined by the spectral dimension and the ``spectral volume" defined in (\ref{vs}), and furthermore that thermodynamical relations will have small oscillatory behavior in the thermodynamic limit. These two properties serve as direct and clear evidence of an underlying fractal structure. To observe such phenomena, we propose studies of thermalization of photons in fractal-shaped (e.g., Sierpinski) mesoscopic waveguides. Of course, one cannot fabricate such a waveguide to all orders of fractal iteration, and the photon wavelength sets a natural cutoff scale. However, computations show that the role of the spectral dimension, and even the oscillatory behavior from the complex poles, can be seen after only about five iterations, which should be accessible. A similar argument appears in \cite{berry}. Another interesting example is provided by systems such as a Fibonacci distribution of dielectric layers \cite{fibonacci}, whose spatial structure is not fractal, but whose energy spectrum is given by a fractal triadic Cantor set. The study of such layered structures would be significant, because here the geometric volume is well defined, and non-fractal; but we predict that the thermodynamically significant volume is the spectral volume.

Acknowledgements: We acknowledge support from the ISF (EA), from the DOE (GD), and from the NSF (AT).

    \end{document}